\documentclass[11pt]{article}
\usepackage{amssymb,latexsym,amsmath,amsbsy}
\usepackage[dvips]{graphicx}
\usepackage{cite}
\usepackage{hyperref}
\newtheorem{theo}{Theorem}

\newtheorem{prop}[theo]{Proposition}
\def\nn{\nonumber}
\def\deg{\mathop{\rm deg}\nolimits}
\def\ch{\mathop{\rm char}\nolimits}
\def\qdots{\mathinner{\mkern1mu\raise1pt\vbox{\kern7pt\hbox{.}}\mkern2mu \raise4pt\hbox{.}\mkern2mu\raise7pt\hbox{.}\mkern1mu}}
\def\Z{{\mathbb Z}}

\def\C{{\mathbb C}}
\def\gl{\mathfrak{gl}}

\def\u{\mathfrak{u}}
\def\h{\mathfrak{h}}
\def\so{\mathfrak{so}}
\def\sp{\mathfrak{sp}}
\def\osp{\mathfrak{osp}}
\def\lb{[\![}
\def\rb{]\!]}
\def\r{\mathsf{r}}

\begin{document}
\begin{center}
{\Large \bf
A class of infinite-dimensional representations of the Lie\\[1mm] 
superalgebra $\osp(2m+1|2n)$ and the parastatistics Fock space } \\[5mm]
{\bf N.I.~Stoilova}\footnote{E-mail: stoilova@inrne.bas.bg}\\[1mm] 
Institute for Nuclear Research and Nuclear Energy,\\ 
Boul.\ Tsarigradsko Chaussee 72, 1784 Sofia, Bulgaria\\[2mm] 
{\bf J.\ Van der Jeugt}\footnote{E-mail: Joris.VanderJeugt@UGent.be}\\[1mm]
Department of Applied Mathematics, Computer Science and Statistics, Ghent University,\\
Krijgslaan 281-S9, B-9000 Gent, Belgium.
\end{center}

%\addtolength{\baselineskip}{2mm}
%\addtolength{\abovedisplayskip}{1mm}
%\addtolength{\belowdisplayskip}{1mm}
%\addtolength{\parskip}{2mm}

\begin{abstract}
An orthogonal basis of weight vectors for a class of infinite-dimensional representations of the orthosymplectic Lie superalgebra $\osp(2m+1|2n)$ is introduced. 
These representations are particular lowest weight representations $V(p)$, with a lowest weight of the form
$[-\frac{p}{2},\ldots, -\frac{p}{2}|\frac{p}{2},\ldots, \frac{p}{2}]$, $p$ being a positive integer.
Explicit expressions for the transformation of the basis under the action of algebra generators are found. 
Since the relations of algebra generators correspond to the defining relations of $m$ pairs of parafermion operators and $n$ pairs of paraboson operators with relative parafermion relations, the parastatistics Fock space of order~$p$ is also explicitly constructed.
Furthermore, the representations $V(p)$ are shown to have interesting characters in terms of supersymmetric Schur functions, and a simple character formula is also obtained. 
\end{abstract}

\vskip 10mm
\noindent Running title: Lie superalgebra $\osp(2m+1|2n)$ and parastatistics


%%%%%%%%%%%%%%%%%%%%%%%%%%%%%%%%%%%%%%%%%%%%%%%%%%%%%%%%%%%%%%%%%%%%%%%%%%%%%%%
%%%%% section
%%%%%%%%%%%%%%%%%%%%%%%%%%%%%%%%%%%%%%%%%%%%%%%%%%%%%%%%%%%%%%%%%%%%%%%%%%%%%%%
\setcounter{equation}{0}
\section{Introduction} \label{sec:Introduction}%

Lie superalgebras and their representation theory have found many applications in physics, especially after the concept of supersymmetry was introduced. Lie superalgebras are relevant in elementary particle physics~\cite{FF}, condensed matter physics~\cite{BGM}, nuclear physics~\cite{BBI}, string theory~\cite{B}, etc.

The finite-dimensional irreducible modules over any basic Lie superalgebra are fully classified~\cite{Kac1} 
and for most of the classes of basic classical Lie superalgebras their characters are known.
However, the construction of explicit matrix elements for all such modules (and especially the so-called indecomposible modules) remains an open problem. 
Partial results giving expressions for matrix elements for various types of $\gl(m|n)$ representations are given in~\cite{T,P1,P2,CGC,M,GIW1,GIW2}. 
In the present paper we make a step further in this respect not for finite-dimensional but for a class of infinite-dimensional irreducible representations of the orthosymplectic Lie superalgebra $\osp(2m+1|2n)$. 
The main reason for considering a particular class of infinite-dimensional irreducible representations of $\osp(2m+1|2n)$ is its connection with the so-called parastatistics Fock space. 

This parastatistics Fock space $V(p)$ (corresponding to parastatistics of order~$p$) is known to correspond to an
infinite-dimensional irreducible lowest weight representation of $\osp(2m+1|2n)$, with lowest weight
$[-\frac{p}{2},\ldots, -\frac{p}{2}|\frac{p}{2},\ldots, \frac{p}{2}]$, $p$ being a positive integer (see Section~4 for more details on this).
When $m=0$, this corresponds to the (infinite-dimensional) paraboson Fock space of $\osp(1|2n)$ with lowest weight $[\frac{p}{2},\ldots, \frac{p}{2}]$, a complete construction of which has been given in~\cite{paraboson}.
When $n=0$, this corresponds to the (finite-dimensional) parafermion Fock space of $\so(2m+1)$ with lowest weight $[-\frac{p}{2},\ldots, -\frac{p}{2}]$, constructed in~\cite{parafermion}. 
Combining the two is far from trivial. 
The subject was already treated in~\cite{NS}, but there only the case $m=n=1$ could be solved.
In this paper, we solve the problem (i.e.\ the construction of $V(p)$ and the explicit action of the generators of
$\osp(2m+1|2n)$ on an appropriate set of basis vectors of $V(p)$) for general $m$ and~$n$.

The paper is organized as follows. 
In Section~2 we establish the notation and give the $\osp(2m+1|2n)$ algebraic structure. Herein, the set of $2m+2n$ generators satisfying certain triple relations are of importance.
Section~3 is devoted to the construction of the infinite-dimensional representations $V(p)$ of $\osp(2m+1|2n)$ and to finding their matrix elements. 
We first describe an induced module construction of a Verma module $\overline V(p)$, of which $V(p)$ is a quotient module. 
For this, the subalgebra chain $\osp(2m+1|2n) \supset \so(2m+1)\oplus\sp(2m) \supset \u(m|n)$ is important, leading to an appropriate labeling of the basis vectors of $\overline V(p)$ in terms of $\u(m|n)$ Gelfan-Zetlin patterns.
This essentially converts the construction of matrix elements to the search for reduced matrix elements of the Lie superalgebra generators, denoted by $G_k$.
Our main computational result is the expression for these reduced matrix elements, given in subsection~3.2. 
These expressions are important, as they yield all matrix elements of $\osp(2m+1|2n)$ generators in this representation.
Furthermore, they allow to identify the basis vectors of the corresponding irreducible representation $V(p)$ itself, and its matrix elements.
The structure of $V(p)$, its character and a character formula are given in subsection~3.3.
In Section~4 we give more details about the important relation between the constructed $\osp(2m+1|2n)$ modules and the parastatistics Fock space. We conclude the paper with some remarks.

As mentioned earlier, the case $m=n=1$ was already treated by one of us in an earlier paper~\cite{NS}.
In that paper~\cite{NS}, the general problem being treated now was already described. 
In order to make the current paper readable, and in particular to fix the notation, we need to repeat some definitions and terminology.
So there is some overlap with~\cite{NS}, but we have tried to keep this minimal.

%%%%%%%%%%%%%%%%%%%%%%%%%%%%%%%%%%%%%%%%%%%%%%%%%%%%%%%%%%%%%%%%%%%%%%%%%%%%%%%
%%%%% section
%%%%%%%%%%%%%%%%%%%%%%%%%%%%%%%%%%%%%%%%%%%%%%%%%%%%%%%%%%%%%%%%%%%%%%%%%%%%%%%
\section{The Lie superalgebras $B(m|n)\equiv \osp(2m+1|2n)$}
\setcounter{equation}{0} \label{sec:B}

The orthosymplectic Lie superalgebra $B(m|n)\equiv \osp(2m+1|2n)$~\cite{Kac} can be defined as the set of all matrices of the form
\begin{equation}
\left(\begin{array}{ccccc} a&b&u&x&x_1 \\
c&-a^t&v&y&y_1\\
-v^t&-u^t&0&z&z_1\\
y_1^t&x_1^t&z_1^t&d&e\\
-y^t&-x^t&-z^t&f&-d^t
\end{array}\right).
\label{osp}
\end{equation}
In~\eqref{osp} $a$ is any $(m\times m)$-matrix, $b$ and $c$ are skew symmetric $(m\times m)$-matrices, 
$u$ and $v$ are $(m\times 1)$-matrices, $x,y,x_1,y_1$ are $(m\times n)$-matrices, $z$ and
$z_1$ are $(1\times n)$-matrices, $d$ is any $(n\times n)$-matrix,
and $e$ and $f$ are symmetric $(n\times n)$-matrices. 
The even (odd) part of $\osp(2m+1|2n)$ is given by all block diagonal (block off-diagonal) matrices in~\eqref{osp}.
Denote by $e_{ij}$ the matrix with zeros everywhere except a $1$ on position $(i,j)$, where the row and column indices run from $1$ to $2m+2n+1$.
Then as a basis in the Cartan subalgebra $\h$ of $\osp(2m+1|2n)$ consider $h_i=e_{ii}-e_{i+m,i+m}$ ($i=1,\ldots,m$), 
$h_{m+j}=e_{2m+1+j,2m+1+j}-e_{2m+1+n+j,2m+1+n+j}$ ($j=1,\ldots,n$). 
In terms of the dual basis $\epsilon_i$ ($i=1,\ldots, m$), $\delta_j$ ($j=1,\ldots,n$) of $\h^*$, the even and odd
root vectors and corresponding roots of $osp(2m+1|2n)$ are given by
\begin{align*}
e_{jk}-e_{k+m,j+m} & \ \leftrightarrow\  \epsilon_j -\epsilon_k, \qquad j\neq k=1,\ldots ,m,\\
e_{j,k+m}-e_{k,j+m} & \ \leftrightarrow\  \epsilon_j +\epsilon_k, \qquad j<k=1,\ldots ,m,\\
e_{j+m,k}-e_{k+m,j} & \ \leftrightarrow\  -\epsilon_j -\epsilon_k, \quad j<k=1,\ldots ,m,\\
e_{j,2m+1}-e_{2m+1,j+m} & \ \leftrightarrow\  \epsilon_j, \qquad\qquad j=1,\ldots ,m,\\
e_{j+m,2m+1}-e_{2m+1,j} & \ \leftrightarrow\   -\epsilon_j , \quad\qquad j=1,\ldots ,m,\\
e_{2m+1+j,2m+1+k}-e_{n+2m+1+k,n+2m+1+j} & \ \leftrightarrow\  \delta_j -\delta_k, 
\qquad j\neq k=1,\ldots ,n,\\
e_{2m+1+j,2m+1+k+n}+e_{2m+1+k,2m+1+j+n} & \ \leftrightarrow\  \delta_j +\delta_k, \qquad j\leq k=1,\ldots ,n,\\
e_{2m+1+n+j,2m+1+k}+e_{2m+1+n+k,2m+1+j} & \ \leftrightarrow\  -\delta_j -\delta_k, \quad j\leq k=1,\ldots ,n,
\end{align*}
and
\begin{align*}
e_{j,2m+1+k}-e_{2m+1+n+k,j+m} & \ \leftrightarrow\  \epsilon_j -\delta_k, 
\qquad j=1,\ldots ,m; \ k=1,\ldots ,n,\\
e_{m+j,2m+1+k}-e_{2m+1+n+k,j} & \ \leftrightarrow\  -\epsilon_j -\delta_k, 
\;\quad j=1,\ldots ,m; \ k=1,\ldots , n,\\
e_{2m+1,2m+1+k}-e_{2m+1+n+k,2m+1} 
& \ \leftrightarrow\  -\delta_k, \;\quad \qquad k=1,\ldots ,n,\\
e_{j,2m+1+n+k}+e_{2m+1+k,m+j} & \ \leftrightarrow\  \epsilon_j +\delta_k, 
\qquad j=1,\ldots ,m; \ k=1,\ldots, n,\\
e_{m+j,2m+1+n+k}+e_{2m+1+k,j} & \ \leftrightarrow\  -\epsilon_j +\delta_k, 
\quad j=1,\ldots ,m; \ k=1,\ldots, n,\\
e_{2m+1,2m+1+n+k}+e_{2m+1+k,2m+1} & \ \leftrightarrow\  \delta_k, \qquad\qquad k=1,\ldots ,n
\end{align*}
respectively.
Introduce the following multiples of the even root vectors with roots 
$\pm \epsilon_j$ ($j=1,\ldots, m$) 
\begin{align}
&c_{j}^+=f_{j}^+= \sqrt{2}(e_{j, 2m+1}-e_{2m+1,j+m}), \nn\\
&c_{j}^-=f_{j}^-= \sqrt{2}(e_{2m+1,j}-e_{j+m,2m+1}), \;
\label{f-as-e}
\end{align}
and of the odd root vectors with roots 
$\pm \delta_j$ ($j=1,\ldots,n$)
\begin{align}
&c_{m+j}^+=b_{j}^+= \sqrt{2}(e_{2m+1,2m+1+n+j}+e_{2m+1+j,2m+1}), \nn\\
&c_{m+j}^-=b_{j}^-= \sqrt{2}(e_{2m+1,2m+1+ j}-e_{2m+1+n+j,2m+1}). \; 
\label{b-as-e}
\end{align}
The operators $c_j^+$ are positive root vectors, and the $c_j^-$ are negative root vectors.
Then the following holds~\cite{Palev1}

\begin{theo}[Palev]
As a Lie superalgebra defined by generators and relations, 
$\osp(2m+1|2n)$ is generated by $2m+2n$ elements $c_j^\pm$ subject to the following relations
\begin{align}
& \lb\lb c_{ j}^{\xi}, c_{ k}^{\eta}\rb , c_{l}^{\epsilon}\rb =-2
\delta_{jl}\delta_{\epsilon, -\xi}\epsilon^{\langle l \rangle} 
(-1)^{\langle k \rangle \langle l \rangle }
c_{k}^{\eta} +2 \epsilon^{\langle l \rangle }
\delta_{kl}\delta_{\epsilon, -\eta}
c_{j}^{\xi}, \label{paraosp}\\
& \qquad\qquad \xi, \eta, \epsilon =\pm\hbox{ or }\pm 1;\quad j,k,l=1,\ldots,n+m. \nn 
\end{align} 
\end{theo}
Herein, $\lb\cdot,\cdot\rb$ is the Lie superalgebra bracket, $\lb a, b \rb = ab-(-1)^{{\deg(a)\deg(b)}}ba$, so it denotes a commutator $[\cdot,\cdot]$ or an anticommutator $\{\cdot,\cdot\}$. Furthermore,
\[
\langle i\rangle \equiv\deg(c_i^\pm)= 
 \left\{ \begin{array}{lll}
 {0} & \hbox{if} & i=1,\ldots ,m \\ 
 {1} & \hbox{if} & i=m+1,\ldots ,n+m.
 \end{array}\right.
\]
The so-called triple relations~\eqref{paraosp} are important: they combine a system of $m$ pairs of parafermion operators with a system of $n$ pairs of paraboson operators in a particular way, see Section~4.

%%%%%%%%%%%%%%%%%%%%%%%%%%%%%%%%%%%%%%%%%%%%%%%%%%%%%%%%%%%%%%%%%%%%%%%%%%%%%%%
%%%%% section
%%%%%%%%%%%%%%%%%%%%%%%%%%%%%%%%%%%%%%%%%%%%%%%%%%%%%%%%%%%%%%%%%%%%%%%%%%%%%%%
\section{Explicit representations of the Lie superalgebra $\osp(2m+1|2n)$}
\setcounter{equation}{0} \label{sec:C}

\subsection{General considerations}

In order to construct a class of representations of $\osp(2m+1|2n)$ we can use an induced module procedure with an appropriate chain of subalgebras. 
This construction was already discussed in~\cite{NS}, and follows the ideas of earlier constructions for $\osp(1|2n)$ and $\so(2m+1)$~\cite{paraboson,parafermion}.
Here, the relevant subalgebras have a simple basis in terms of the given generators:
\begin{prop}
A basis for the even subalgebra $\so(2m+1)\oplus \sp(2n)$ of $\osp(2m+1|2n)$ is given by the elements
\begin{align}
&[c_i^\xi, c_k^\eta],\quad c_l^\epsilon,\qquad (i,k,l=1,\ldots,m;\ \xi,\eta,\epsilon =\pm) \nn \\
& \{c_{m+j}^\xi, c_{m+s}^\eta\}, \qquad(j,s=1,\ldots,n;\ \xi,\eta =\pm).
\label{sp2n}
\end{align}
The $(m+n)^2$ elements 
\begin{equation}
\lb c_j^+, c_k^-\rb \qquad(j,k=1,\ldots,m+n)
\label{un}
\end{equation} 
are a basis for the subalgebra $\u(m|n)$.
\end{prop}
Note that, due to the triple relations~\eqref{paraosp} 
\begin{equation}
[c_i^-,c_i^+]=-2 h_i\quad (i=1,\ldots,m), \qquad \{c_{m+j}^-,c_{m+j}^+\}=2 h_{m+j}\quad (j=1,\ldots,n).
\label{Cartan-h}
\end{equation}
Hence $\h=\hbox{span}\{h_i,\ i=1,\ldots,m+n\}$, the Cartan subalgebra of $\osp(2m+1|2n)$, is also the Cartan subalgebra of $\so(2m+1)\oplus \sp(2n)$ and of $\u(m|n)$.
This allows us to define a one-dimensional $\u(m|n)$ module $\C |0\rangle$, spanned on a vector $|0\rangle$, by setting
$\lb c_j^-,c_k^+\rb |0\rangle = p\,\delta_{jk}\, |0\rangle$, 
where $p$ is a positive integer.
So this is just a trivial $\u(m|n)$ module with one weight vector of weight 
$[-\frac{p}{2},\ldots, -\frac{p}{2}|\frac{p}{2},\ldots, \frac{p}{2}]$.
The reason for considering this particular weight will become clear later.

The subalgebra $\u(m|n)$ can be extended to a parabolic subalgebra ${\mathfrak P}$ of $\osp(2m+1|2n)$
\begin{equation}
{\mathfrak P} = \hbox{span} \{ c_j^-, \lb c_j^+, c_k^-\rb, \lb c_j^-, c_k^-\rb \;|\;
j,k=1,\ldots,m+n \},
\label{P}
\end{equation}
and by requiring $c_j^- |0\rangle =0$ ($j=1,\ldots,m+n$), the one-dimensional module 
$\C |0\rangle$ is extended to a one-dimensional ${\mathfrak P}$ module.
Now we define the induced $\osp(2m+1|2n)$ module $\overline V(p)$ 
\begin{equation}
 \overline V(p) = \hbox{Ind}_{\mathfrak P}^{\osp(2m+1|2n)} \C|0\rangle.
 \label{defInd}
\end{equation}
This is an $\osp(2m+1|2n)$ representation with lowest weight 
$[-\frac{p}{2},\ldots, -\frac{p}{2}|\frac{p}{2},\ldots, \frac{p}{2}]$.
By the Poincar\'e-Birkhoff-Witt theorem~\cite{Kac1}, it is easy to write a basis for $\overline V(p)$:
\begin{align}
& (c_1^+)^{k_1}\cdots (c_{m+n}^+)^{k_{m+n}} (\lb c_1^+,c_2^+\rb)^{k_{12}} (\lb c_1^+,c_3^+\rb)^{k_{13}} \cdots 
(\lb c_{m+n-1}^+,c_{m+n}^+\rb)^{k_{m+n-1,m+n}} |0\rangle, \label{Vpbasis}\\
& \qquad k_1,\ldots,k_{m+n},k_{12},k_{13}\ldots,k_{m-1,m},k_{m+1,m+2},k_{m+1,m+3}\ldots,k_{m+n-1,m+n} \in \Z_+, \nn\\
& \qquad k_{1,m+1},k_{1,m+2}\ldots,k_{1,m+n},k_{2,m+1},\ldots,k_{m,m+n} \in\{0,1\}. \nn
\end{align}
However in general $\overline V(p)$ is not an irreducible module of
$\osp(2m+1|2n)$. If $M(p)$ is the maximal nontrivial submodule of $\overline V(p)$ then the
irreducible module is
\begin{equation}
V(p) = \overline V(p) / M(p).
\label{Vp}
\end{equation}
The purpose is now to determine the vectors belonging to $M(p)$ and also to find explicit matrix elements of
the $\osp(2m+1|2n)$ generators $c_j^\pm$ in an appropriate basis of $V(p)=\overline V(p)/M(p)$. 

{}From the basis~\eqref{Vpbasis} of $\overline V(p)$, and the fact that the character of $\C |0\rangle$ is just given by 
\[
(x_1)^{-p/2}\cdots (x_m)^{-p/2}(y_1)^{p/2}\cdots (y_n)^{p/2},
\]
it follows that the character of $\overline V(p)$ as a formal series of terms 
$\nu x_1^{j_1}x_2^{j_2}\ldots x_m^{j_m} y_1^{j_{m+1}}y_2^{j_{m+2}}\ldots y_n^{j_{m+n}}$, 
with $[j_1,\ldots,j_m|j_{m+1},\ldots,j_{m+n}]$ a weight of $\overline V(p)$
and $\nu$ the dimension of this weight space,
is given by: 
\begin{equation}
 \ch \overline V(p) = \frac{(x_1)^{-p/2}\cdots (x_m)^{-p/2}(y_1)^{p/2}\cdots (y_n)^{p/2} \prod_{i,j}(1+x_iy_j)}
{\prod_{i}(1-x_i)\prod_{i<k}(1-x_ix_k)\prod_{j}(1-y_j)\prod_{j<l}(1-y_jy_l)} .
 \label{char-osp}
\end{equation}
This expression has an expansion in terms of supersymmetric Schur functions,
valid for general $m$ and $n$. 
The origin of this expansion goes back to unpublished work of Cummins and King~\cite{Cummins,CumminsKing}, 
see also~\cite[Corollary~3.4]{Loday}.

Throughout the paper, we let ${\cal{H}}$  denote the set of all partitions $\lambda$ satisfying the so-called
$(m|n)$-hook condition $\lambda_{m+1}\leq n$ \cite{Mac}.
\begin{prop}{\bf(Cummins, King)}
Let $({\mathbf x})=(x_1, x_2,\ldots, x_m)$ 
and $({\mathbf y})=(y_1, y_2,\ldots, y_n)$ 
be sets of $m$ and $n$ variables, respectively. Then
\begin{equation}
\frac{\prod_{i,j}(1+x_iy_j)}{\prod_{i}(1-x_i)\prod_{i<k}(1-x_ix_k)\prod_{j}(1-y_j)\prod_{j<l}(1-y_jy_l)} = 
\sum_{\lambda \in {\cal{H}}} s_\lambda (x_1,\ldots,x_m|y_1,\ldots,y_n)
= \sum_{\lambda \in {\cal{H}}} s_\lambda ({\mathbf x}|{\mathbf y}). 
\label{Schur}
\end{equation}
In the right hand side, 
$s_\lambda({\mathbf x}|{\mathbf y})$ is the supersymmetric Schur function~\cite{Mac} defined by
\[
s_\lambda({\mathbf x| \mathbf y})= \sum_{\tau} s_{\lambda/\tau}({\mathbf x})s_{\tau'}({\mathbf y})
= \sum_{\sigma, \tau} c_{\sigma \tau}^\lambda s_{\sigma}({\mathbf x})s_{\tau'}({\mathbf y}),
\] 
with $\ell(\sigma)\leq m$; \; $\ell(\tau')\leq n$; \;$\tau'$ the conjugate partition to $\tau$;\; $c_{\sigma \tau}^\lambda$ the Littlewood-Richardson coefficients;
$|\lambda|= |\sigma|+|\tau|$ and $s_{\nu}({\mathbf x})$ 
the ordinary Schur function. 
(As usual, $\ell(\sigma)$ and $|\sigma|$ denote the length and weight of the partition $\sigma$, see~\cite{Mac} for the common notation regarding partitions and symmetric functions.)
\end{prop} 

The objects appearing in the right hand side of~\eqref{Schur}, i.e.\ the supersymmetric Schur functions $s_\lambda(x|y)$ with 
$\lambda \in {\cal H}$, are precisely the characters of the irreducible covariant $\u(m|n)$ tensor representations $V([\Lambda^\lambda])$. 
The relation between the partitions $\lambda=(\lambda_1, \lambda_2,\ldots), \; \lambda_{m+1}\leq n$
and the highest weights $\Lambda^\lambda$ $\equiv [\mu ]^r\equiv [\mu_{1r},\ldots,
\mu_{mr}|\mu_{m+1,r}\ldots ,\mu_{rr}]$; $r=m+n$, of the irreducible covariant $\u(m|n)$ tensor representations
 is given by~\cite{JHKR}: 
\begin{align}
& \mu_{ir}=\lambda_i, \quad 1\leq i\leq m, \label{hwpart1}\\
& \mu_{m+i,r}=\max\{0, \lambda'_i-m\}, \quad 1\leq i\leq n, \label{hwpart2}
\end{align}
where $\lambda'$ is the partition conjugate~\cite{Mac} to $\lambda$. 
In such a way the expansion~(\ref{Schur}) yields the branching to $\u(m|n)$ of the $\osp(2m+1|2n)$
representation $\overline V(p)$ and a possibility to label the basis vectors of $\overline V(p)$.
For each irreducible covariant $\u(m|n)$ tensor representations one can use the corresponding
Gelfand-Zetlin basis (GZ)~\cite{CGC}. The union of all these GZ-bases is then the basis for $\overline V(p)$.
Thus the basis of $\overline V(p)$ consists of vectors of the form ($p$ is dropped from the notation of the vectors)

\begin{equation}
|\mu)\equiv |\mu)^r = \left|
\begin{array}{lclllcll}
\mu_{1r} & \cdots & \mu_{m-1,r} & \mu_{mr} & \mu_{m+1,r} & \cdots & \mu_{r-1,r}
& \mu_{rr}\\
\mu_{1,r-1} & \cdots & \mu_{m-1,r-1} & \mu_{m,r-1} & \mu_{m+1,r-1} & \cdots
& \mu_{r-1,r-1} & \\
\vdots & \vdots &\vdots &\vdots & \vdots & \qdots & & \\
\mu_{1,m+1} & \cdots & \mu_{m-1,m+1} & \mu_{m,m+1} & \mu_{m+1,m+1} & & & \\
\mu_{1m} & \cdots & \mu_{m-1,m} & \mu_{mm} & & & & \\
\mu_{1,m-1} & \cdots & \mu_{m-1,m-1} & & & & & \\
\vdots & \qdots & & & & & & \\
\mu_{11} & & & & & & &
\end{array}
\right)
= \left| \begin{array}{l} [\mu]^r \\[2mm] |\mu)^{r-1} \end{array} \right),
\label{mn}
\end{equation}
which satisfy the conditions~\cite{CGC}
\begin{equation}
 \begin{array}{rl}
1. & \mu_{ir}\in{\mathbb Z}_+ \; \hbox{are fixed and } \mu_{jr}-\mu_{j+1,r}\in{\mathbb Z}_+ , \;j\neq m,\;
 1\leq j\leq r-1,\\
 & \mu_{mr}\geq \# \{i:\mu_{ir}>0,\; m+1\leq i \leq r\};\\
2.& \mu_{is}-\mu_{i,s-1}\equiv\theta_{i,s-1}\in\{0,1\},\quad 1\leq i\leq m;\;
 m+1\leq s\leq r;\\
3. & \mu_{ms}\geq \# \{i:\mu_{is}>0,\; m+1\leq i \leq s\}, \quad m+1\leq s\leq r ;\\ 
4.& \hbox{if }\;
\mu_{m,m+1}=0, \hbox{then}\; \theta_{mm}=0; \\
5.& \mu_{is}-\mu_{i+1,s}\in{\mathbb Z}_+,\quad 1\leq i\leq m-1;\;
 m+1\leq s\leq r-1;\\
6.& \mu_{i,j+1}-\mu_{ij}\in{\mathbb Z}_+\hbox{ and }\mu_{i,j}-\mu_{i+1,j+1}\in{\mathbb Z}_+,\\
 & 1\leq i\leq j\leq m-1\hbox{ or } m+1\leq i\leq j\leq r-1.
 \end{array}
\label{cond3}
\end{equation}
The weight of $|0\rangle$ is $[-\frac{p}{2},\ldots,-\frac{p}{2}|\frac{p}{2},\ldots, \frac{p}{2}]$; therefore the weight
of the vector $|\mu)$ is determined by
\begin{equation}
h_{k}|\mu)=\left(\mp \frac{p}{2}+\sum_{j=1}^k \mu_{jk}-\sum_{j=1}^{k-1} \mu_{j,k-1}\right)|\mu),
\label{hkm}
\end{equation}
where the minus sign is taken when $k=1,\ldots,m$ and the plus sign when $k=m+1,\ldots,m+n$.
{}From the triple relations~\eqref{paraosp}, it follows that
under the $\u(m|n)$ basis~\eqref{un}, the set $(c_1^+, c_2^+,\ldots,c_{r}^+)$ is a standard $\u(m|n)$ tensor of rank (1,0,\ldots,0). This means that to every $c_j^+$ one can attach 
 a unique GZ-pattern with top line
$1 0\ldots 0$:
\begin{equation}
c_j^+ \sim \begin{array}{l}1 0 \cdots 0 0 0\\[-1mm]
1 0 \cdots 0 0\\[-1mm] \cdots \\[-1mm] 0 \cdots 0\\[-1mm] \cdots\\[-1mm] 0 \end{array},
\label{cGZ}
\end{equation}
where the pattern consists of $j-1$ zero rows at the bottom, and the first $r-j+1$ rows are of the form
$1 0 \cdots 0$.
The tensor product rule in $\u(m|n)$ reads
\begin{equation}
([\mu]^r) \otimes (1 0\cdots 0) = ([\mu]^r_{+1}) \oplus ([\mu]^r_{+2}) \oplus \cdots \oplus([\mu]^r_{+r})
\label{umntensor}
\end{equation}
where $([\mu]^r) = (\mu_{1r},\mu_{2r},\ldots,\mu_{rr})$ and a subscript $\pm k$ indicates an increase of 
the $k$th label by $\pm 1$:
\begin{equation}
([\mu]^r_{\pm k}) = (\mu_{1r},\ldots,\mu_{kr}\pm 1,\ldots, \mu_{rr}).
\label{mu+k}
\end{equation}
In the right hand side of~\eqref{umntensor}, only those components which satisfy conditions 1.  in~\eqref{cond3} survive.
The matrix elements of $c_j^+$ can now be written as follows:
\begin{align}
(\mu' | c_j^+ | \mu ) & = 
\left( \begin{array}{ll} [\mu]^r_{+k} \\[1mm] |\mu')^{r-1} \end{array} \right| c_j^+
\left| \begin{array}{ll} [\mu]^r \\[1mm] |\mu)^{r-1} \end{array} \right) \nn\\
& = \left( \begin{array}{ll} [\mu]^r \\[2mm] |\mu)^{r-1} \end{array} ; \right.
 \begin{array}{l}1 0 \cdots 0 0\\[-1mm]
1 0 \cdots 0\\[-1mm] \cdots\\[-1mm] 0 \end{array} 
\left| \begin{array}{ll} [\mu]^r_{+k} \\[2mm] |\mu')^{r-1} \end{array} \right)
\times
([\mu]^n_{+k}||c^+||[\mu]^r).
\label{mmatrix}
\end{align}
The first factor in the right hand side of~\eqref{mmatrix} is a $\u(m|n)$ Clebsch-Gordan coefficient (CGC) given by formulae
(4.9)-(4.17) in~\cite{CGC}, 
and the second factor is a {\em reduced matrix element}.
The possible values of the patterns $\mu'$ are determined by the $\u(m|n)$ tensor product rule and the first line of $\mu'$
is of the form~\eqref{mu+k}. 
The purpose is now to find expressions for the reduced matrix elements:
\begin{equation}
G_k([\mu]^r) = G_k(\mu_{1r},\mu_{2r},\ldots,\mu_{rr}) = ([\mu]^r_{+k}||c^+||[\mu]^r),
\label{Gk}
\end{equation}
for arbitrary $r$-tuples $[\mu]^r=(\mu_{1r},\mu_{2r},\ldots,\mu_{rr})$ that correspond to highest weights.
If these can be determined, then one has in a way obtained explicit actions of the Lie superalgebra generators $c_j^\pm$ on a basis of $\overline V(p)$:
\begin{align}
c_j^+|\mu) & = \sum_{k,\mu'} \left( \begin{array}{ll} [\mu]^r \\[2mm] |\mu)^{r-1} \end{array}\right. ;
 \begin{array}{l}1 0 \cdots 0 0\\[-1mm]
1 0 \cdots 0\\[-1mm] \cdots\\[-1mm] 0 \end{array} 
\left| \begin{array}{ll} [\mu]^r_{+k} \\[2mm] |\mu')^{r-1} \end{array} \right)
G_k([\mu]^r) \left| \begin{array}{ll} [\mu]^r_{+k} \\[1mm] |\mu')^{r-1} \end{array} \right), \label{cj+r}\\
c_j^-|\mu) & = \sum_{k,\mu'} \left( \begin{array}{ll} [\mu]_{-k}^r \\[2mm] |\mu')^{r-1} \end{array}\right. ;
 \begin{array}{l}1 0 \cdots 0 0\\[-1mm]
1 0 \cdots 0\\[-1mm] \cdots\\[-1mm] 0 \end{array} 
\left| \begin{array}{ll} [\mu]^r \\[2mm] |\mu)^{r-1} \end{array} \right)
G_k([\mu]_{-k}^r) \left| \begin{array}{ll} [\mu]^r_{-k} \\[1mm] |\mu')^{r-1} \end{array} \right). \label{cj-r}
\end{align}
So all that is left is the determination of the unknown functions $G_k$.

\subsection{Computation of the expressions $G_k$}

We decided to collect some of the technical aspects of this computation in a separate subsection.
In order to determine the unknown functions $G_k$ we started from the following action:
\begin{equation}
\{ c_r^-, c_r^+ \} |\mu) = 2h_r |\mu) = 
\Bigl(p+2(\sum_{j=1}^r \mu_{jr}-\sum_{j=1}^{r-1}\mu_{j,r-1} )\Bigr) |\mu), \label{crcr}
\end{equation}
We can express the left-hand side by means of~\eqref{cj+r} and~\eqref{cj-r}, applying the explicit formulae for the 
CGCs. The result is a very complicated system of recurrence relations for the functions $G_k$.
To have some idea of the complexity, let us present the relation resulting from the diagonal action in the left hand side of~\eqref{crcr} explicitly:
\begin{align}
&\sum_{i=1}^m\left((1-\theta_{i,r-1})   
\prod_{j\neq i=1}^m \left(\frac{\mu_{ir}-\mu_{jr}-i+j+1}{\mu_{ir}-\mu_{j,r-1}-i+j}\right)  
 \frac{\prod_{s=m+1}^{r-1}  
(\mu_{ir}+\mu_{s,r-1}+2m-i-s+1 )}
{ \prod_{s=m+1}^{r} (\mu_{ir}+\mu_{sr}+2m-i-s+2)}G_i^2(\mu)\right) \nn\\[1mm]
+&\sum_{i=1}^m\left(\theta_{i,r-1}   
\prod_{j\neq i=1}^m \left(\frac{\mu_{ir}-\mu_{jr}-i+j}{\mu_{ir}-\mu_{j,r-1}-i+j-1}\right)  
 \frac{\prod_{s=m+1}^{r-1}  
(\mu_{ir}+\mu_{s,r-1}+2m-i-s )}
{ \prod_{s=m+1}^{r} (\mu_{ir}+\mu_{sr}+2m-i-s+1)}G_i^2(\mu)_{-ir}\right) \nn\\[1mm]
+&\sum_{q=m+1}^r \left(  
\prod_{j=1}^m \left(\frac{\mu_{jr}+\mu_{qr}+2m-j-q+1}{\mu_{j,r-1}+\mu_{q,r}+2m-j-q+2}\right)  
 \frac{\prod_{s=m+1}^{r-1}  
(\mu_{qr}-\mu_{s,r-1}-q+s+1 )}
{ \prod_{s\neq q=m+1}^{r} (\mu_{qr}-\mu_{sr}-q+s)}G_q^2(\mu)\right) \nn\\[1mm]
+&\sum_{q=m+1}^r \left(  
\prod_{j=1}^m \left(\frac{\mu_{jr}+\mu_{qr}+2m-j-q}{\mu_{j,r-1}+\mu_{q,r}+2m-j-q+1}\right)  
 \frac{\prod_{s=m+1}^{r-1}  
(\mu_{qr}-\mu_{s,r-1}-q+s )}
{ \prod_{s\neq q=m+1}^{r} (\mu_{qr}-\mu_{sr}-q+s-1)}G_q^2(\mu)_{-qr}\right) \nn\\[1mm]
=& p + 2\left(\sum_{i=1}^r\mu_{ir}-\sum_{i=1}^{r-1}\mu_{i,r-1}\right). \label{id2}
\end{align}
Herein, 
\[
G_k(\mu)\equiv  G_k(\mu_{1r}, \mu_{2r},\ldots, \mu_{rr}); \quad 
G_k(\mu)_{-ir}\equiv  G_k(\mu_{1r},\ldots \mu_{i-1,r}, \mu_{ir}-1, \mu_{i+1,r}\ldots, \mu_{rr}),
\]
and the $\theta$-values in~\eqref{id2} take values in $\{0,1\}$ in correspondence with the conditions~\eqref{cond3}.

Using the relevant boundary conditions and all possible $\theta$-values, we have been able to solve this 
system of recurrence relations. This task would have been hardly impossible without the use of Maple. 
It turns out that the relations~\eqref{id2} alone are already sufficient to yield a solution for the squares of the unknown functions, $G_k^2$. 
Then, it is a matter of using the relations following from the non-diagonal terms in the left hand side of~\eqref{crcr}
in order to determine the signs of $G_k$.
Finally, the last step is to verify that with the solution thus obtained, all triple relations~\eqref{paraosp} acting on $|\mu)$ are satisfied. This sounds like a tremendous task to perform; however, it is the simplest step of the computation.

Without going into the details of these computational steps, let us present the final result.
\begin{prop}
\label{prop-main}
The reduced matrix elements $G_k$ ($k=1,\ldots, m+n=r$) appearing in the actions of $c_j^\pm$ on vectors
$|\mu)$ of $\overline V(p)$ are given by:
\begin{align}
G_{k}(\mu_{1r}, \mu_{2r},\ldots, \mu_{rr}) &=
\left(-\frac{
({\cal E}_m(\mu_{kr}+m-n-k)+1)\prod_{j\neq k=1}^{m} (\mu_{kr}-\mu_{jr}-k+j)}
{\prod_{j\neq \frac{k}{2}=1}^{\lfloor m/2 \rfloor} (\mu_{kr}-\mu_{2j,r}-k+2j)
(\mu_{kr}-\mu_{2j,r}-k+2j+1)}
\right)^{1/2} \nn\\
& \times \prod_{j=1}^n\left(\frac{
\mu_{kr}+\mu_{m+j,r}+m-j-k+2}
{\mu_{kr}+\mu_{m+j,r}+m-j-k+2-{\cal E}_{m+\mu_{m+j,r}}}
\right)^{1/2} \label{Gkeven}
\end{align}
for $k\leq m$ and $k$ even; 
\begin{align}
& G_{k}(\mu_{1r}, \mu_{2r},\ldots, \mu_{rr}) 
=\nn\\
&
\left(\frac{(p-\mu_{kr}+k-1)
({\cal O}_m(\mu_{kr}+m-n-k)+1)\prod_{j\neq k=1}^{m} (\mu_{kr}-\mu_{jr}-k+j)}
{\prod_{j\neq \frac{k+1}{2}=1}^{\lceil m/2 \rceil} (\mu_{kr}-\mu_{2j-1,r}-k+2j-1)
(\mu_{kr}-\mu_{2j-1,r}-k+2j)}
\right)^{1/2} \nn\\
& \times \prod_{j=1}^n\left(\frac{
\mu_{kr}+\mu_{m+j,r}+m-j-k+2}
{\mu_{kr}+\mu_{m+j,r}+m-j-k+2-{\cal O}_{m+\mu_{m+j,r}}}
\right)^{1/2}
\label{Gkodd}
\end{align}
for $k\leq m$ and $k$ odd. The remaining expressions are
\begin{align}
& G_{m+k}(\mu_{1r}, \mu_{2r},\ldots, \mu_{rr}) 
=(-1)^{\mu_{m+k+1,r}+\mu_{m+k+2,r}+\ldots+\mu_{rr}}\nn\\
&
\times \left(
({\cal O}_{\mu_{m+k,r}}(\mu_{m+k,r}-k+n)+1)({\cal E}_{m+\mu_{m+k,r}}(p+\mu_{m+k,r}+m-k)+1)
\right)^{1/2} \nn\\
&
\times \left(\frac{
\prod_{j=1}^{\lfloor m/2 \rfloor} ({\cal E}_{m+\mu_{m+k,r}}(\mu_{2j,r}+\mu_{m+k,r}-2j-k+m+1)+1)}
{\prod_{j=1}^{\lceil m/2 \rceil} ({\cal E}_{m+\mu_{m+k,r}}(\mu_{2j-1,r}+\mu_{m+k,r}-2j-k+m+1)+1)}\right)^{1/2}
 \nn\\
&
\times \left(\frac{
\prod_{j=1}^{\lceil m/2 \rceil} ({\cal O}_{m+\mu_{m+k,r}}(\mu_{2j-1,r}+\mu_{m+k,r}-2j-k+m+2)+1)}
{\prod_{j=1}^{\lfloor m/2 \rfloor} ({\cal O}_{m+\mu_{m+k,r}}(\mu_{2j,r}+\mu_{kr}-2j-k+m)+1)}\right)^{1/2}\nn\\
& \times \prod_{j\neq k=1}^n\left(\frac{
\mu_{m+j,r}-\mu_{m+k,r}-j+k}
{\mu_{m+j,r}-\mu_{m+k,r}-j+k-{\cal O}_{\mu_{m+j,r}-\mu_{m+k,r}}}
\right)^{1/2}
\label{Gm+k}
\end{align}
for $k=1,2,\ldots,n$.
\end{prop}
Herein ${\cal E}$ and ${\cal O}$ are the even and odd functions defined by
\begin{align}
& {\cal E}_{j}=1 \hbox{ if } j \hbox{ is even and 0 otherwise},\nn\\
& {\cal O}_{j}=1 \hbox{ if } j \hbox{ is odd and 0 otherwise}; \label{EO}
\end{align}
where obviously ${\cal O}_j=1-{\cal E}_j$, but it is still convenient to use both notations.
Also, note that products such as $\prod_{j\neq k=1}^{s}$ means ``the product over all $j$-values
running from 1 to $s$, but excluding $j=k$''. The notation $\lfloor a \rfloor$ (resp.\ $\lceil a \rceil$)
refers to the {\em floor} (resp.\ {\em ceiling}) of
$a$, i.e.\ the largest integer not exceeding~$a$ (resp.\ the smallest integer greater than or equal to $a$).

\subsection{Structure of $V(p)$ and its character}

Although the actual computation of the reduced matrix elements $G_k$ was quite involved, the final result presented in Proposition~\ref{prop-main} is in fact fairly simple.
These general expressions of the matrix elements in $\overline V(p)$ allow the determination of the structure of the irreducible representations $V(p)$. 
Indeed, taking into account the general conditions~\eqref{cond3}, 
the only factor in the right hand sides of~\eqref{Gkeven}-\eqref{Gm+k} that may become zero appears in~\eqref{Gkodd} and is
\[
p-\mu_{kr}+k-1 \qquad (k\leq m \ \hbox{and}\ k\ \hbox{odd}).
\]
In particular for $k=1$ this factor is $(p-\mu_{1r})$, and $\mu_{1r}$ is the largest
integer in the first row of the GZ-pattern~\eqref{mn} 
(which is also the first part of the partition $\lambda$, see~\eqref{hwpart1}).
Starting from the vacuum vector, with a GZ-pattern 
consisting of all zeros, one can raise the entries in the GZ-pattern by applying
the operators $c_j^+$. However, when $\mu_{1r}$ has reached the value~$p$ it can no
longer be increased. As a consequence, all vectors $|\mu)$ with $\mu_{1r}>p$ belong to
the submodule $M(p)$. This uncovers the structure of $V(p)$.

\begin{theo}
\label{prop-main2}
An orthonormal basis for the space $V(p)$ is given by the vectors $|\mu)$,
see~\eqref{mn}-\eqref{cond3}, with $\mu_{1r}\leq p$.
The action of the Cartan algebra elements of $\osp(2m+1|2n)$ is:
\begin{align}
& h_{k}|\mu)=\left(- \frac{p}{2}+\sum_{j=1}^k \mu_{jk}-\sum_{j=1}^{k-1} \mu_{j,k-1}\right)|\mu),\quad k=1,\ldots,m;\nn \\
& h_{k}|\mu)=\left (\frac{p}{2}+\sum_{j=1}^k \mu_{jk}-\sum_{j=1}^{k-1} \mu_{j,k-1}\right)|\mu),\quad k=m+1,\ldots,r.
\end{align}
The action of the operators $c_j^\pm, \; j=1,\ldots,r$ is given by~\eqref{cj+r}-\eqref{cj-r}, where the CGCs are found in~\cite{CGC} (see formulae (4.9)-(4.17)) and the reduced matrix elements are given by~\eqref{Gkeven}-\eqref{Gm+k}.
\end{theo}

The condition $\mu_{1r}\leq p$ is equivalent to $\lambda_1\leq p$, see~\eqref{hwpart1}, so the partition $\lambda$ should have width less than or equal to $p$.
Following~\eqref{char-osp} and~\eqref{Schur}, we have

\begin{theo}
\label{prop-char1}
The character of the irreducible representation $V(p)$ is given by
\begin{equation}
\ch V(p) = {\mathbf x}^{-p/2}{\mathbf y}^{p/2} 
\sum_{\lambda \in {\cal{H}},\; \lambda_1\leq p} s_\lambda ({\mathbf x}|{\mathbf y}). 
\label{charVp1}
\end{equation}
Herein, ${\mathbf x}^{-p/2}{\mathbf y}^{p/2}$ is the short hand notation for $(x_1)^{-p/2}\cdots (x_m)^{-p/2}(y_1)^{p/2}\cdots (y_n)^{p/2}$, and the sum is over all partitions $\lambda$ satisfying the $(m|n)$-hook condition ($\lambda_{m+1}\leq n$) and having width less than or equal to $p$ ($\lambda_1\leq p$).
\end{theo}

Eq.~\eqref{charVp1} gives the character of $V(p)$ as an (infinite) sum of supersymmetric Schur functions. 
It would be interesting to have an actual character formula, as in~\eqref{char-osp}, with denominator factors corresponding to the even positive roots and numerator factors corresponding to odd positive roots.
For this purpose, we need to recall some further notation for partitions~\cite{Mac}.
For a general partition $\lambda=(\lambda_1,\lambda_2,\ldots)$, the Young diagram $F^\lambda$ consists of $|\lambda|$ boxes
arranged in $\ell(\lambda)$ left-adjusted rows of length $\lambda_i$. 
If $F^\lambda$ has $r$ boxes on the main diagonal with ``arm length'' $a_k$ and ``leg length'' $b_k$ (i.e.\ there are $a_k$ boxes just to the right of the $k$th diagonal box, and $b_k$ boxes just below this diagonal box), then $\lambda$ is said to have rank $\r(\lambda)=\r$ and in Frobenius notation one writes
\[
\lambda = \left( \begin{array}{c} a_1\ a_2\ \cdots \ a_\r \\ b_1\ b_2\ \cdots \ b_\r \end{array}\right),
\]
where $a_1>a_2>\cdots >a_\r\geq 0$ and $b_1>b_2>\cdots >b_\r\geq 0$.
Note that in this notation, the conjugate of $\lambda$ is just
\[
\lambda' = \left( \begin{array}{c} b_1\ b_2\ \cdots \ b_\r \\ a_1\ a_2\ \cdots \ a_\r \end{array}\right).
\]
There are some special families of partitions. Usually, ${\mathcal P}$ denotes the set of all partitions, including also the zero partition $(0)$. Another important family is
\[
{\mathcal P}_0 = \{ \lambda = \left( \begin{array}{c} a_1\ a_2\ \cdots \ a_\r \\ a_1\ a_2\ \cdots \ a_\r \end{array}\right),
\quad (\r=0,1,2,\ldots)\},
\]
i.e.\ the set of all self-conjugate partitions (satisfying $\lambda'=\lambda$).
For a positive integer $p$, the set of partitions that will be of importance for us is~\cite[(6)]{King2013}
\begin{equation}
{\mathcal P}_p = \{ \lambda = \left( \begin{array}{c} a_1\ a_2\ \cdots \ a_\r \\ b_1\ b_2\ \cdots \ b_\r \end{array}\right),
\quad a_k=b_k+p\ (k=1,2,\ldots,\r), \quad (\r=0,1,2,\ldots)\}.
\label{Pp}
\end{equation}
In particular, let us denote by ${\cal H}_p$ the set of partitions $\lambda$ of the form~\eqref{Pp} satisfying the $(m|n)$-hook condition:
\[
{\cal H}_p = {\cal H} \bigcap {\mathcal P}_p.
\]
Now we have:

\begin{theo}
\label{prop-char2}
A character formula for the irreducible representation $V(p)$ is given by
\begin{equation}
\ch V(p) = {\mathbf x}^{-p/2}{\mathbf y}^{p/2} 
\frac{\prod_{i,j}(1+x_iy_j)\sum_{\sigma \in {\cal H}_p} (-1)^{(|\sigma|-\r(\sigma)(p-1))/2}s_\sigma ({\mathbf x}|{\mathbf y})}{\prod_{i}(1-x_i)\prod_{i<k}(1-x_ix_k)\prod_{j}(1-y_j)\prod_{j<l}(1-y_jy_l)}. 
\label{charVp2}
\end{equation}
\end{theo}

\noindent {\bf Proof.} 
According to~\eqref{Schur} and~\eqref{charVp1}, we only need to prove that
\begin{equation}
\sum_{\lambda \in {\cal{H}},\; \lambda_1\leq p} s_\lambda ({\mathbf x}|{\mathbf y}) =
\sum_{\nu\in{\cal H}} s_\nu({\mathbf x}|{\mathbf y}) 
\sum_{\sigma \in {\cal H}_p} (-1)^{(|\sigma|-\r(\sigma)(p-1))/2}s_\sigma ({\mathbf x}|{\mathbf y}).
\label{supers1}
\end{equation}
For such type of identities, it is common to work first with an infinite number of variables $x_1, x_2,\ldots$ and $y_1, y_2,\ldots$, establish the identity, and finally restrict to a finite number of variables~\cite{Mac}.
When an ordinary Schur function $s_\lambda$ is restricted to a finite number of variables $x_1, x_2,\ldots,x_m$,
one has $s_\lambda({\mathbf x})=0$ whenever $\ell(\lambda)>m$.
Similarly, when a supersymmetric Schur functions $s_\lambda ({\mathbf x}|{\mathbf y})$ is restricted to a finite number of variables $x_1, x_2,\ldots,x_m$, $y_1, y_2,\ldots,y_n$, one has $s_\lambda ({\mathbf x}|{\mathbf y})=0$ if $\lambda$ does not satisfy the $(m|n)$-hook condition. Hence, it is sufficient to prove the universal identity (i.e.\ with an infinite number of variables for ${\mathbf x}$ and ${\mathbf y}$)
\begin{equation}
\sum_{\lambda \in {\mathcal P},\; \lambda_1\leq p} s_\lambda ({\mathbf x}|{\mathbf y}) =
\sum_{\nu\in{\mathcal P}} s_\nu({\mathbf x}|{\mathbf y}) 
\sum_{\sigma \in {\mathcal P}_p} (-1)^{(|\sigma|-\r(\sigma)(p-1))/2}s_\sigma ({\mathbf x}|{\mathbf y}).
\label{supers2}
\end{equation}
Now one can work out the right hand side of~\eqref{supers2} using the product rule for supersymmetric Schur functions.
It is well known~\cite{Berele,King1990} that this is governed by the Littlewood-Richardson coefficients: in other words, the product rule for supersymmetric Schur functions $s_\lambda ({\mathbf x}|{\mathbf y})$ is just the same as that for ordinary Schur functions $s_\lambda ({\mathbf x})$.
But for ordinary Schur functions the right hand side is well known by the identity~\cite[(28)]{King2013}
\begin{equation}
\sum_{\nu\in{\mathcal P}} s_\nu({\mathbf x}) 
\sum_{\sigma \in {\mathcal P}_p} (-1)^{(|\sigma|-\r(\sigma)(p-1))/2}s_\sigma ({\mathbf x}) =
\sum_{\lambda \in {\mathcal P},\; \lambda_1\leq p} s_\lambda ({\mathbf x}).
\label{s1}
\end{equation}
Therefore, the right hand side of~\eqref{supers2} becomes
\[
\sum_{\lambda \in {\mathcal P},\; \lambda_1\leq p} s_\lambda ({\mathbf x}|{\mathbf y})
\]
and this finishes the proof.

Note that an alternative (more involved) proof of the identity~\eqref{supers1} is also contained in~\cite{Loday}.

%%%%%%%%%%%%%%%%%%%%%%%%%%%%%%%%%%%%%%%%%%%%%%%%%%%%%%%%%%%%%%%%%%%%%%%%%%%%%%%
%%%%% section
%%%%%%%%%%%%%%%%%%%%%%%%%%%%%%%%%%%%%%%%%%%%%%%%%%%%%%%%%%%%%%%%%%%%%%%%%%%%%%%
\section{The parastatistics Fock space}
\setcounter{equation}{0} \label{sec:D}
The motivation for the present paper comes from some physical ideas, and more precisely from some conceptual difficulties of quantum mechanics.  In 1950 Wigner posed the question: Do the equations of motion determine the
quantum mechanical commutation relations?~\cite{Wigner}. On the simplest example, the one dimensional harmonic oscillator, he showed that a more general approach is to start from the equations of motion, Hamilton's equations and the Heisenberg equations, instead of assuming that the position and momentum operators are subject to the canonical commutation relations. Actually he found an infinite set of solutions of the compatibility conditions of Hamilton's equations and the Heisenberg equations and among them is the solution with the canonical commutation relations. Each of these solutions is now known to correspond to a representation of $\osp(1|2)$. Wigner's solutions were generalized in 1953 by Green  who wrote down the so called paraboson relations~\cite{Green}.  Paraboson operators are generalizations of Bose operators. He also generalized the ordinary Fermi operators to parafermion operators~\cite{Green}. Paraoperators were applied to quantum field theory~\cite{Haag, Wu, Ohnuki}, generalizations of quantum statistics~\cite{Green, GM, Tolstoy2014, YJ, YJ2, KA, KK, KD} (parastatistics), and in Wigner quantum systems~\cite{Palev86, Palev82, Palev92}.
Parafermion operators $f_j^\pm$ ($j=1,\ldots,m$), satisfying~\cite{Green}
\begin{equation}
[[f_{ j}^{\xi}, f_{ k}^{\eta}], f_{l}^{\epsilon}]=\frac 1 2
(\epsilon -\eta)^2
\delta_{kl} f_{j}^{\xi} -\frac 1 2 (\epsilon -\xi)^2
\delta_{jl}f_{k}^{\eta}, 
\label{f-rels}
\end{equation}
where $j,k,l\in \{1,2,\ldots,m\}$ and $\eta, \epsilon, \xi \in\{+,-\}$ (to be interpreted as $+1$ and $-1$
in the algebraic expressions $\epsilon -\xi$ and $\epsilon -\eta$), generate 
the orthogonal Lie algebra $\mathfrak{so}(2m+1)$~\cite{Kamefuchi,Ryan} and 
$n$ pairs of parabosons $b_j^\pm$~\cite{Green}, satisfying
\begin{equation}
[\{ b_{ j}^{\xi}, b_{ k}^{\eta}\} , b_{l}^{\epsilon}]= (\epsilon -\xi) \delta_{jl} b_{k}^{\eta} 
 + (\epsilon -\eta) \delta_{kl}b_{j}^{\xi}, 
\label{b-rels}
\end{equation}
generate the orthosymplectic Lie superalgebra 
$\mathfrak{osp}(1|2n)$~\cite{Ganchev}. 
The important objects, 
parafermion and paraboson Fock spaces are characterized by a parameter $p$, and their explicit construction was given recently in~\cite{parafermion} (for parafermions) and in~\cite{paraboson} (for parabosons). The problem of natural extension to a combined system of $m$ pairs of parafermions and $n$ pairs of parabosons with explicit construction of the corresponding Fock space was so far solved only for $m=n=1$~\cite{NS}. 

The commutation relations among para-operators were studied in~\cite{GM} (see also~\cite{Tolstoy2014} for a summary).
The result is that for each pair of para-operators there can exist at most four types of relative commutation relations: straight commutation, straight anticommutation, relative paraboson, and relative parafermion relations. 
The case with relative paraboson relations and the corresponding Fock representations has been investigated in~\cite{YJ,YJ2,KA,KK,KD}. 
It was proved by Palev~\cite{Palev1} that $m$ 
parafermions $f_j^\pm\equiv c_j^\pm$ and $n$ parabosons $b_j^\pm\equiv c_{m+j}^\pm$ 
with relative parafermion relations~\eqref{paraosp}
 generate the
orthosymplectic Lie superalgebra $\mathfrak{osp}(2m+1|2n)$. 
By definition the parastatistics Fock space $V(p)$ is the Hilbert space with vacuum vector $|0\rangle$, 
defined by means of ($j,k=1,2,\ldots,m+n$)
\begin{align}
& \langle 0|0\rangle=1, \qquad c_j^- |0\rangle = 0, \qquad (c_j^\pm)^\dagger = c_j^\mp,\nn\\
& \lb c_j^-,c_k^+ \rb |0\rangle = p\delta_{jk}\, |0\rangle, \label{Fock}
\end{align}
and by irreducibility under the action of the algebra spanned by
the elements $c_j^+$, $c_j^-$, $j=1,\ldots,m+n$, subject
to~\eqref{paraosp}. 
The parameter $p$ is referred to as the order of the parastatistics system
 and for $p=1$ the parastatistics Fock
space $V(p)$ coincides with the Fock space of $n$ bosons and $m$ fermions with unusual grading 
which anticommute~\cite{Palev1}. 
Therefore the constructed infinite-dimensional irreducible $\osp(2m+1|2n)$ modules $V(p)$ are 
precisely the parastatistics Fock spaces.

%%%%%%%%%%%%%%%%%%%%%%%%%%%%%%%%%%%%%%%%%%%%%%%%%%%%%%%%%%%%%%%%%%%%%%%%%%%%%%
%%%%% section
%%%%%%%%%%%%%%%%%%%%%%%%%%%%%%%%%%%%%%%%%%%%%%%%%%%%%%%%%%%%%%%%%%%%%%%%%%%%%%%
\setcounter{equation}{0}
\section{Summary and conclusion} \label{sec:summary}

Obtaining explicit matrix elements for a class of Lie (super)algebras and a particular class of irreducible representations is always a very hard problem.
One instance where this was successfully solved was for $\gl(m|n)$, where the actions of the Chevalley generators were computed for the class of covariant representations (in an appropriate GZ-basis)~\cite{CGC}.

Stimulated by a problem that had been open for many years, we considered at the time a particular class of representations for the Lie superalgebras $\osp(1|2n)$ in~\cite{paraboson}. 
These representations $V(p)$ of $\osp(1|2n)$ have a specific lowest weight of the form $(\frac{p}{2}, \ldots, \frac{p}{2})$, with $p$ a positive integer, and they are infinite dimensional. Explicit matrix elements were obtained for a set of generators of $\osp(1|2n)$. These are not the Chevalley generators, but the paraboson generators of $\osp(1|2n)$.
This solved the long standing problem of constructing the so-called paraboson Fock space of order~$p$.

In a similar fashion, the construction of the parafermion Fock space was tackled.
The para\-fermion Fock spaces $V(p)$ (of order~$p$) correspond to a class of finite-dimensional highest weight representations of the Lie algebra $\so(2m+1)$, with highest weight $(\frac{p}{2}, \ldots, \frac{p}{2})$ and lowest weight $(-\frac{p}{2}, \ldots, -\frac{p}{2})$. Here again, an appropriate basis was constructed, and matrix elements of the (parafermion) generators of $\so(2m+1)$ were obtained~\cite{parafermion}.
In fact, the results were even generalized for an infinite number of paraboson or parafermion generators~\cite{infinite}.

The main challenge was then to consider a combined system of parabosons and parafermions, a so-called parastatistics Fock space. 
From an algebraic point of view, the most interesting type of parastatistics occurs when $m$ pairs of parafermions and $n$ pairs of parabosons are combined according to relative parafermion relations, since this corresponds to a set of generators of the Lie superalgebra $\osp(2m+1|2n)$~\cite{Palev1}.
The parastatistics Fock space $V(p)$ then corresponds to an infinite-dimensional irreducible lowest weight representation of $\osp(2m+1|2n)$, with lowest weight $[-\frac{p}{2},\ldots, -\frac{p}{2}|\frac{p}{2},\ldots, \frac{p}{2}]$. 
This is the class of representations considered in this paper.

The objective of this paper was then to find explicit matrix elements for the parastatistics generators of $\osp(2m+1|2n)$ for this class of representations. 
Although the methods of the previous papers~\cite{paraboson,parafermion} -- based on an induced module construction and an appropriate GZ-basis -- could be used, the computational problem turned out to be much harder than expected.
This computational problem consists of finding the solution of recurrence relations like~\eqref{id2}.
We consider it as a big achievement that we managed to solve this problem.

The method used here gives also a description of the branching of the $\osp(2m+1|2n)$ representation $V(p)$ to $\u(m|n)$ (or $\gl(m|n)$).
In other words, the character of $V(p)$ is easy to give in terms of supersymmetric Schur functions.
This allowed us to give also an attractive character formula for $V(p)$, see Theorem~\ref{prop-char2}, which we consider as the second main result of this paper.

The current paper deals only with one type of parastatistics, where the relative commutation relation between parafermions and parabosons is of type~\eqref{paraosp}, according to the so-called relative parafermion relations.
For other types of relative relations, the algebraic structure generated by the para-operators was so far less obvious.
In a recent paper, Tolstoy~\cite{Tolstoy2014} showed that parastatistics with relative paraboson relations is also connected to a $\Z_2\times \Z_2$-graded Lie superalgebra. 
This result could be the starting point for the study of representations of such $\Z_2\times \Z_2$-graded Lie superalgebras. 
Next step could be investigation of the eventual connection of parastatistics to the affine Lie (super)algebras and their modules.

\section*{Acknowledgments}
The authors were supported by Joint Research Project ``Representation theory of Lie (super)al\-ge\-bras and generalized quantum statistic'' in the framework of an international collaboration programme between the Research Foundation – Flanders (FWO) and the Bulgarian Academy of Sciences. NIS was partially supported by Bulgarian NSF grant DFNI T02/6.

\end{document}